\documentclass[12pt]{report}

\usepackage[a4paper, left=3cm, right=3cm, top=3cm, bottom=3cm]{geometry}
\usepackage{setspace}
\usepackage{amsmath, amssymb, amsfonts}
\usepackage{graphicx}
\usepackage{xcolor}
\usepackage{empheq}
\usepackage[]{hyperref}

\begin{document}
\renewcommand{\thesection}{}
\renewcommand{\thesubsection}{}
\noindent \textbf{\Large \textsc{Emergent tuning heterogeneity in cortical circuits is sensitive to cellular neuronal dynamics}} \\

\vspace{4mm}

\noindent \textbf{\textsc{Mohammadreza Soltanipour\textsuperscript{1,2}, Stefan Treue\textsuperscript{3} \& Fred Wolf\textsuperscript{1,2}}} \\

\vspace{2mm}

\noindent \textsuperscript{1}\small{G{\"o}ttingen Campus Institute for Dynamics of Biological Networks, University of Göttingen, G{\"o}ttingen, Germany} \\
\noindent \textsuperscript{2}\small{Max Planck Institute for Dynamics and Self-Organization, G{\"o}ttingen, Germany} \\
\noindent \textsuperscript{3}\small{Cognitive Neuroscience Laboratory, German Primate Center, G{\"o}ttingen, Germany} 

\begin{center}
    \textbf{Abstract}
\end{center}
\begin{quote}
Cortical circuits exhibit high levels of response diversity, even across apparently uniform neuronal populations. While emerging data-driven approaches exploit this heterogeneity to infer effective models of cortical circuit computation (e.g. Genkin et al. Nature 2025), the power of response diversity to enable inference of mechanistic circuit models is largely unexplored. Within the landscape of cortical circuit models, spiking neuron networks in the balanced state naturally exhibit high levels of response and tuning diversity emerging from their internal dynamics. A statistical theory for this emergent tuning heterogeneity, however, has only been formulated for binary spin models (Vreeswijk $\&$ Sompolinsky, 2005). 
Here we present a formulation of feature-tuned balanced state networks that allows for arbitrary and diverse dynamics of postsynaptic currents and variable levels of heterogeneity in cellular excitability but nevertheless is analytically exactly tractable with respect to the emergent tuning curve heterogeneity. Using this framework, we present a case study demonstrating that, for a wide range of parameters even the population mean response is non-universal and sensitive to mechanistic circuit details. As our theory enables exactly and analytically obtaining the likelihood-function of tuning heterogeneity given circuit parameters, we argue that it forms a powerful and rigorous basis for neural circuit inference.
\end{quote}

\section{Introduction}
Ring models are widely used in neuroscience to understand how neural circuits represent continuous variables and sharpen weakly tuned inputs. The seminal work of Ben-Yishai, Bar-Or, and Sompolinsky demonstrated how recurrent excitation and inhibition on a ring can amplify thalamic input to produce sharply selective orientation responses in primary visual cortex \cite{ben1995theory}. Similar architectures have been extended to spatial working memory, where recurrent excitation stabilizes persistent activity bumps \cite{compte2000synaptic, wang2001synaptic, vafidis2022learning}, to decision making, where ring-like recurrent dynamics support ramping activity and winner-take-all computations \cite{wong2006recurrent}, and to head-direction circuits, where a localized bump of activity encodes the animal’s angular orientation \cite{zhang1996representation, taube2007head, noorman2024maintaining}. The biological importance of this motif was further underscored by the discovery of an anatomically realized ring of neurons in the fruit fly visual system, directly supporting ring-like architectures as substrates of orientation selectivity \cite{seelig2013feature, kim2017ring, biswas2024fly}.

A distinct class of models, classical balanced networks, provide a framework for understanding
how irregular spiking at the single-neuron level can coexist with reliable and stable population
responses \cite{van1996chaos,van1998chaotic,wolf2014dynamical}. When synaptic weights scale as $1/\sqrt{K}$, with $K$ the average number of inputs
per neuron, excitation and inhibition dynamically cancel to leading order, leaving temporal
fluctuations of order one. Since the connectivity is sparse, these residual fluctuations are
effectively independent across neurons, resulting in irregular and weakly correlated spiking
activity, producing cortical-like activity patterns \cite{ecker2010decorrelated}. This mechanism
gives rise to broad firing-rate distributions even under constant external input and deterministic,
identical neuronal dynamics, and has been confirmed to hold also in densely connected networks
\cite{renart2010asynchronous}.

Vreeswijk and Sompolinsky introduced balanced ring models that combine the ring architecture
with balanced-state dynamics \cite{van2005irregular}. Unlike classical ring attractors, where
sharpening arises through recurrent amplification, balanced rings achieve sharpening primarily
via inhibitory feedback that suppresses untuned components of the activity. This distinct
mechanism highlights how inhibitory stabilization can also give rise to selective tuning,
consistent with later work showing that orientation selectivity in primary visual cortex can be
explained within the balanced network framework even in the absence of structured orientation
selective connectivity \cite{hansel2012mechanism}. Extensions of the balanced ring framework
have incorporated short-term synaptic plasticity, which can generate bistability in random
networks \cite{mongillo2012bistability} and account for irregular persistent activity in ring-based
models of working memory \cite{hansel2013short}.

In the large-$K$ limit, the balance equation enforces cancellation of average excitatory and
inhibitory inputs, ensuring a finite net drive per neuron. Although this condition constrains
population activity, it has two key limitations. First, it fixes only the first moment of activity
and does not constrain higher moments, thereby failing to account for heterogeneity and the
broad firing-rate distributions observed in cortical activity. Second, even at the level of the
mean, it may not uniquely determine activity in networks with low-rank structure (e.g., ring
networks with cosine-modulated connectivity), where additional self-consistency conditions are
required. Mean-field self-consistency methods have been developed only for binary neuron
models \cite{van2005irregular}, whereas studies of spatially structured balanced networks have
so far focused on the balance equation alone \cite{rosenbaum2014balanced}.

Here we build on the superior tractability of the Gauss–Rice neuron model to introduce and
analyze a balanced ring network that can be studied in closed form. This approach enables us
to go beyond the balance equation and derive self-consistent solutions for networks with
structured connectivity. Within this framework, we investigate how inhibitory stabilization
sharpens input tuning, how heterogeneity emerges across neurons, and under which conditions
the balanced ring remains stable. In doing so, we bridge the conceptual gap between classical
balanced ring networks and the cortical heterogeneity observed in the spiking activity of
cortical neurons.

In addition to the Gauss–Rice framework, we also considered a balanced network with von
Mises–modulated connectivity. In this case, the balance equation determines the mean activity
exactly, and by appropriately tuning its parameters, the von Mises kernel can be matched to
the cosine-modulated connectivity of the ring model. This provides a direct way to compute
the population mean response without invoking self-consistency methods or specifying a
particular neuron model. Using this approach, we tested whether the population mean
response is a universal feature of balanced ring architectures, independent of the neuronal
transfer function, while the Gauss–Rice analysis is required to capture heterogeneity and
firing rate distribution of the network.

\section{Results}
\subsection{Gauss--Rice neuron model}

To map Gaussian input currents onto firing rates, we employ the Gauss--Rice (GR) neuron model
\cite{schmidt2023analytically}. Its dynamics are identical to a leaky integrate--and--fire neuron
except that the membrane potential does not reset after a spike. The voltage $V(t)$ evolves as
\begin{equation}
\tau_M \dot V = -V + I(t),
\end{equation}
where $\tau_M$ is the membrane time constant and $I(t)$ is the total synaptic input, including
both external and recurrent contributions. A spike is emitted whenever $V(t)$ crosses a fixed
threshold $\Psi_0$ with positive slope. Since both $V(t)$ and $\dot V(t)$ are Gaussian random
processes, the firing rate can be evaluated using Rice’s formula for the expected rate of
upward threshold crossings.

The resulting input--output relation is \cite{tchumatchenko2010correlations}:
\begin{equation}
\boxed{
\begin{aligned}
&\nu(I,\sigma_V,\tau_S) = \nu_m \exp\!\left[-\frac{(I-\Psi_0)^2}{2\sigma_V^2}\right], \\[6pt]
&\sigma_V^2 = C_{V}(0) \qquad \sigma_{\dot{V}} ^2 = -C^{\prime\prime}_{V}(0) \\
\nu_m &= \frac{\sigma_{\dot{V}}}{2\pi\sigma_{V}} = \frac{1}{2\pi \tau_{S}} \qquad 
\tau_{S} = \sqrt{\frac{C_{V}(0)}{-C^{\prime\prime}_{V}(0)}} 
\end{aligned}}
\label{eq:fi_curve_gauss0}
\end{equation}

Where $C_{V}(\tau)$ is the autocovariance 
function of the voltage fluctuations, 
and $\tau_{S}$ is their differential correlation time. 
Since $C_{V}(\tau)$ can be calculated for 
any type of input current statistics, once 
the current covariance 
function $C_{I}(\tau)$ is known, 
the Gauss–Rice neuron has 
an analytically tractable noisy f–I curve 
for arbitrary synaptic receptor complement, 
in the validity range of the 
diffusion approximation.  

For the minimal case of an exponentially 
decaying current correlation function with 
correlation time $\tau_{I}$ one obtains  

\begin{equation}
\boxed{
\begin{aligned}
&\nu(I,\sigma_I,\tau_I) = \nu_m \exp\!\left[-\frac{(I-\Psi_0)^2}{2\sigma_V^2}\right], \\[6pt]
&\sigma_V^2 = \frac{\sigma_I^2 \tau_I}{\tau_I+\tau_M} 
\qquad \nu_m = \frac{1}{2\pi\sqrt{\tau_I \tau_M}}.
\end{aligned}}
\label{eq:fi_curve_gauss}
\end{equation}

which is used in the following for concreteness. Here $I$ is the mean input current, $\sigma_I^2$ its temporal variance, $\tau_I$ its temporal correlation time, and $\sigma_V^2$ the variance of the membrane potential. The firing rate thus takes the
form of a Gaussian function of the mean input, with width $\sigma_V^2$ set by the temporal
fluctuations. The maximal rate $\nu_m$ depends only on $\tau_I$ and
$\tau_M$. Together, these relations define a compact and analytically tractable transfer function, making the Gauss--Rice model ideally suited for mean-field descriptions of balanced networks.

\subsection{Random balanced networks with Gauss--Rice neurons}

We begin with a random inhibitory network consisting of $N$ neurons. Connectivity is given by
an Erd\H{o}s--R\'enyi graph in which each ordered pair $(i,j)$ forms a synapse independently
with probability $K/N$, so that each neuron receives on average $K$ inputs. Synaptic weights
are $J_{ij} = -J_{0}/\sqrt{K}$, and each spike generates a postsynaptic current (PSC) with kernel
$f(t)$ normalized such that $\int f(t)\,dt = 1$.

\subsubsection{Compound spike train statistics}

The temporal sequence of spikes from a single presynaptic neuron $j$ constitutes a point process
with instantaneous rate $\nu_j(t)$. The entirety of incoming spikes to a postsynaptic neuron $i$
forms the compound spike train, with rate
\begin{equation}
\Omega_i(t) = \sum_{j \in \text{pre}(i)} \nu_j(t),
\label{eq:compound_random}
\end{equation}
where $\text{pre}(i)$ denotes the set of presynaptic partners of neuron $i$. In the sparse limit
($N \gg K$), presynaptic spike trains are approximately independent, so the compound input
can be treated as a Poisson-like process. By the central limit theorem, for large $K$ the total
synaptic input is Gaussian and fully characterized by its mean and variance,
\begin{equation}
\begin{aligned}
\Omega &= [\Omega_i]_i = K \, \bar\nu, \\
\mathrm{Var}(\Omega) &= [(\delta \Omega_i)^2]_i = K q ,
\end{aligned}
\label{eq:random_compound}
\end{equation}
where $\bar\nu = [\nu_i]_i$ is the mean firing rate and $q = [\nu_i^2]_i$ is the second moment of
the firing-rate distribution. We use $\left[. \right]$ for the network average.

Intuitively, this variance has two contributions: (i) heterogeneity of firing rates across neurons
and (ii) variability of the in-degrees $K_i$. If all neurons had identical firing rate $\bar\nu$ but
random in-degrees, the variance would be $\bar\nu^2 K$. Conversely, if all neurons had the same
in-degree but variable rates, the variance would be $K\,\mathrm{Var}(\nu)$. In general both effects
coexist and add, yielding the compact expression above. A detailed derivation is carried out in \cite{schmidt2023analytically}.

\subsubsection{Statistics of the input current}

Because each neuron receives the superposition of many independent presynaptic spike trains,
the total synaptic input current $I_i(t)$ can be approximated as a Gaussian random process by
the central limit theorem. Such a process is fully characterized by its mean and correlation
function, or equivalently its variance. The input current can be written as
\begin{equation}
I_i(t) = \langle I_i(t) \rangle + \sigma_{I,i} \, \eta_i(t),
\end{equation}
where $\eta_i(t)$ is a Gaussian process of zero mean and unit variance, and
$\sigma_{I,i}^2 = C_{I,i}(0)$ is the temporal variance, defined by the correlation function
\begin{equation}
C_{I,i}(t,t') = \langle \delta I_i(t)\,\delta I_i(t') \rangle .
\end{equation}

For a general postsynaptic current (PSC) kernel $f(t)$ normalized to $\int_0^\infty f(t)\,dt=1$,
the ensemble-averaged input current is
\begin{equation}
\langle I_i(t) \rangle = J \int_0^\infty ds \, f(t-s)\, \Omega_i(s) = J \,\Omega_i .
\end{equation}
The last step is due to the stationary condition, where $\Omega_i(t) \equiv \Omega_i$ is constant. 
The ensemble mean of the input current is equal to its temporal mean,
$I_i := \langle I_i(t) \rangle = \langle I_i(t) \rangle_t$. 
Following Eq.~\ref{eq:random_compound}, the mean and variance of the mean input currents read
\begin{equation}
\begin{aligned}
I_0 & := \big[\langle I_i(t) \rangle\big]_i = J K \bar\nu, \\
\alpha^2 &= \big[(I_i - I_0)^2\big]_i = J^2 K q,
\end{aligned}
\end{equation}
where $q = [\nu_i^2]_i$ is the second moment of the firing-rate distribution.  

Temporal fluctuations of $I_i(t)$ are described by the correlation function
\begin{equation}
C_{I,i}(\Delta t) = J^2 \Omega_i \int_{-\infty}^\infty ds \, f(s) f(s+\Delta t),
\label{eq:crr_compound_spikes}
\end{equation}
which depends only on the time lag $\Delta t$ for stationary input. At zero lag this yields
the temporal variance
\begin{equation}
\sigma_{I,i}^2 = C_{I,i}(0) = J^2 \Omega_i \, C ,
\qquad
C = \int_{-\infty}^\infty ds \, f(s)^2 .
\end{equation}
The constant $C$ depends only on the PSC kernel $f(t)$. 

Since $J_{ij} = -\frac{J_0}{\sqrt{K}}$, the full input current to neuron $i$ can therefore be expressed as
\begin{equation}
\begin{aligned}
I_i(t) &= \sqrt{K}\,(I_{\mathrm{ext}} - J_0 \bar\nu) + \alpha x_i + \sigma_I \eta_i(t), \\
\alpha^2 &= J_0^2 q, \\
\sigma_{I}^2 &= \sigma_{I,i}^2 = J_0^2 C \,\bar\nu .
\end{aligned}
\label{eq:general_input_current}
\end{equation}
Here the first term ensures the balance of external and recurrent input at order $\sqrt{K}$,
the second term $\alpha x_i$ captures quenched heterogeneity across neurons, and the third term
$\sigma_I \eta_i(t)$ represents temporal fluctuations with variance $\sigma_I^2$. Importantly,
in the large-$K$ limit the temporal variance is almost identical for all neurons,
$\sigma_{I,i}^2 \simeq J_0^2 C \bar\nu$.

From the first term we obtain the balance equation,
\begin{equation}
\bar\nu = \frac{I_{\mathrm{ext}}}{J_0}.
\label{eq:balance_eq_random_net}
\end{equation}

\subsubsection{Exponentially decaying PSC}

In the following we specialize to an exponential postsynaptic current (PSC) kernel with decay
time constant $\tau_I$, normalized as
\begin{equation}
f(t) = \frac{1}{\tau_I} \exp\!\left(-\frac{t}{\tau_I}\right) \Theta(t),
\end{equation}
where $\Theta(t)$ is the Heaviside step function. This choice allows us to evaluate the input
correlation function in closed form. Following Eq.~\ref{eq:crr_compound_spikes} we obtain
\begin{equation}
\begin{aligned}
C_{I,i}(t) &= \frac{J^2 \Omega_i}{2 \tau_I} \exp\!\left(-\frac{|t|}{\tau_I}\right), \\
\sigma_{I,i}^2 &= C_{I,i}(0) = \frac{J^2}{2 \tau_I} \, \Omega_i .
\end{aligned}
\label{eq:exp_synapse_sigma}
\end{equation}

With our choice of $J_{ij} = -J_{0}/\sqrt{K}$, the temporal variance becomes
\begin{equation}
\sigma_{I}^2 = \frac{J_{0}^2}{2 \tau_I} \, \bar\nu .
\label{eq:temporal_fluc_random_net}
\end{equation}

\subsubsection{Self-consistency equations}

We now close the loop by combining the Gauss--Rice transfer function with the distribution of
input currents in the random balanced network. The mean rate $\bar\nu$ is fixed by the external
drive through the balance equation
$\bar\nu = \frac{I_{\mathrm{ext}}}{J_0}$,
while the second moment $q = [\nu_i^2]_i$ and the mean input $I_0$ must be determined
self-consistently.

The mean input currents $I_i$ are Gaussian distributed across the population,
\begin{equation}
\rho(I_i) = \frac{1}{\sqrt{2\pi\alpha^2}}
\exp\!\left[-\frac{(I_i-I_0)^2}{2\alpha^2}\right],
\qquad \alpha^2 = J_0^2 q,
\label{eq:gaus_dis_input_current}
\end{equation}
with variance set by the quenched heterogeneity. Each neuron transforms its input via the
Gauss--Rice transfer function (Eq.\ref{eq:fi_curve_gauss},\ref{eq:temporal_fluc_random_net}),
\begin{equation}
\nu(I_i) = \nu_m \exp\!\left[-\frac{(I_i-\Psi_0)^2}{2\sigma_V^2}\right],
\end{equation}
where $\nu_m = (2\pi\sqrt{\tau_I\tau_M})^{-1}$ and
$\sigma_V^2 = J_0^2 \bar\nu / \tau_q$ with $\tau_q = 2(\tau_I+\tau_M)$.

Averaging over the Gaussian distribution of inputs yields the mean rate
\begin{equation}
\bar\nu(q,I_0) = \nu_m \frac{\sigma_V}{\sqrt{\alpha^2+\sigma_V^2}}
\exp\!\left[-\frac{(I_0-\Psi_0)^2}{2(\alpha^2+\sigma_V^2)}\right],
\label{eq:self_mean_compact}
\end{equation}
and the second moment
\begin{equation}
q(q,I_0) = \nu_m^2 \frac{\sigma_V}{\sqrt{2\alpha^2+\sigma_V^2}}
\exp\!\left[-\frac{(I_0-\Psi_0)^2}{2\alpha^2+\sigma_V^2}\right].
\label{eq:self_second_compact}
\end{equation}

Equations~\ref{eq:self_mean_compact} and \ref{eq:self_second_compact} together with the
balance equation \ref{eq:balance_eq_random_net} define the self-consistency equations of the random balanced
Gauss--Rice network, uniquely determining the firing rate distribution of the network.


\subsection{Cosine modulated ring balanced networks with Gauss--Rice neurons}

\begin{figure}[h!]
  \centering
  \includegraphics[width=\textwidth]{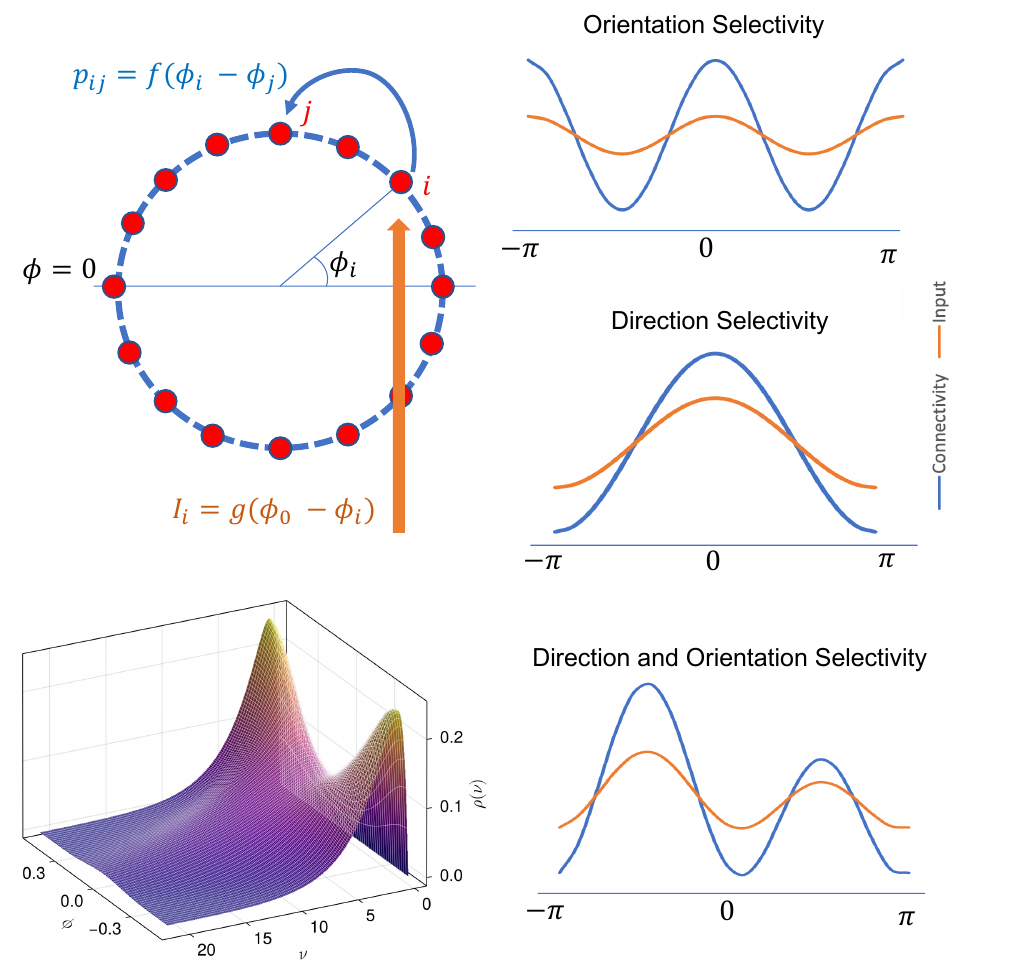} 
  \caption{\textbf{Schematic of ring model and input selectivity.} 
(Left) Neurons are arranged on a ring according to their preferred angle $\phi_i$. 
Connection probability is determined by the angular distance, $p_{ij} = f(\phi_i - \phi_j)$ (blue), and external input by $I_i = g(\phi_0 - \phi_i)$ (orange), centered around stimulus angle $\phi_0$ (without loss of generality $\phi_0 = 0$ is fixed in the analysis). 
(Right) Example input and connectivity kernels. 
For orientation selectivity, only the zeroth and second Fourier modes are required, resulting in $\pi$-periodic tuning.  
For direction selectivity, one additionally needs the first Fourier mode, which breaks the $\pi$-symmetry and yields $2\pi$-periodic responses. A ring model operating in the balanced state will generate emergent response heterogeneity as illustrated by the angle-dependent firing rate distribution shown on the bottom left. The theory presented here enables obtaining it by direct analytical calculation.}
  \label{fig:schematic}
\end{figure}

In our framework, the connection probability between neurons is described by a function 
$f(\phi_i - \phi_j)$ that depends only on the angular distance between 
their preferred orientations (Fig. \ref{fig:schematic}). Expanding $f$ into Fourier components reveals that 
different selectivity regimes can be captured by retaining specific modes. 
For orientation selectivity, it is sufficient to keep the zeroth and second 
Fourier modes, which yield a $\pi$-periodic tuning profile (Fig. \ref{fig:schematic}). 

\newpage

Each neuron $i$ is assigned a preferred orientation $\phi_{i} \in [-\pi/2, \pi/2]$, and external stimulation is centered at orientation $\phi_{0}$. The connection probability between neurons depends on their difference in preferred orientation,
\begin{equation} \label{connection_p_cosine}
P_{ij} = \frac{K}{N}\left[1 + 2p_{c}\cos 2(\phi_{i} - \phi_{j})\right],
\end{equation}
where $p_{c}$ determines the strength of orientation-specific modulation around the homogeneous baseline $K/N$. The total synaptic input current to neuron $i$ is given by
\begin{equation} \label{total_input_cosine}
I_{i}(t) = \sqrt{K}\left[I_{0c} + I_{\mu c}\left(1 + \mu_{c}\cos 2(\phi_{0} - \phi_{i})\right)\right] + \sum_{j \in \mathrm{pre}(i)} J_{ij} f(t - t_{j}^{s}) ,
\end{equation}

Here $\mu_{c}$ modulates the degree of orientation selectivity. The cosine factor ensures that neurons with preferred orientations close to the stimulus $\phi_{0}$ receive stronger input. Without loss of generality we assume $\phi_{0} = 0$.

The last term in Eq.~\ref{total_input_cosine} represents recurrent inhibitory input from presynaptic neurons $j \in \mathrm{pre}(i)$. Synaptic strength is $J_{ij} = -J_0/\sqrt{K}$, while $f(t - t_{j}^{s})$ denotes the postsynaptic current evoked by a presynaptic spike at time $t_{j}^{s}$. This recurrent term implements inhibitory stabilization and provides the mechanism for sharpening untuned components of the activity. 

Following the same reasoning as in the randomly connected network (Eq.~\ref{eq:compound_random}), the total presynaptic spike input to neuron $i$ in the stationary state and in the limit $1 \ll K \ll N$, can be approximated as a compound Poisson process. The effective input rate to neuron $i$ with preferred orientation $\phi_{i}$ then reads
\begin{equation}
\Omega_{i}(\phi_{i}) = \sum_{j \in \mathrm{pre}(i)} \nu_{j}(\phi_{j}) ,
\end{equation}
where $\nu_{j}(\phi_{j})$ denotes the firing rate of presynaptic neuron $j$ with preferred orientation $\phi_{j}$.

Due to the large number of inputs, the central limit theorem applies to the statistics of $\Omega_{i}$. Therefore it can be fully described by its mean and variance:

\begin{equation} \label{eq:mean_varaince_input_rate_ring}
\begin{aligned}
  \left[ \Omega_i(\phi_i) \right]_i &=  \int_{-\pi/2}^{\pi/2} \! \omega(\phi' - \phi)\nu(\phi') \, \mathrm{d}\phi'\\
Var(\Omega) &=\left[ (\Omega_{i}(\phi_{i}) - [\Omega(\phi)])^2 \right]_i =  \int_{-\pi/2}^{\pi/2} \! \omega(\phi' - \phi)\nu^2(\phi') \, \mathrm{d}\phi' \equiv q(\phi)
\end{aligned}
\end{equation} 

 Here $q(\phi)$ is defined as $\int \! \omega(\phi' - \phi)\nu^2(\phi') \, \mathrm{d}\phi'$, and $\omega(\Delta\phi) = \tfrac{K}{\pi}(1 + 2p_{c}\cos 2\Delta\phi)$ is the kernel density function of connection between neurons with $\Delta \phi$ difference in their preferred orientation.

Generalizing the argument in the previous section, the input to neuron $i$ in the network reads
\begin{equation} \label{general_form_input}
I_{i}(t, \phi_{i}) = I_{c}(\phi) + \alpha(\phi)x_{i} + \sigma_{I}(\phi)\,\eta(t),
\end{equation} 
where $I_{c}(\phi)$ is the deterministic input based on the neuron’s location on the ring,
$\alpha(\phi)$ represents quenched disorder, and $\sigma_{I}(\phi)$ captures temporal
fluctuations. 

Using the input rate we can now derive the input current to each neuron from recurrent connections with the time-averaged input:

\begin{equation} \label{average_recurernt_inhibitory}
  I_{rec,i}(\phi_{i}) = \bigg \langle \sum_{j \in pre(i)} \frac{-J_{0}}{\sqrt{K}} f(t - t_{j}^s) \bigg \rangle  = \frac{-J_{0}}{\sqrt{K}} \Omega_{i}(\phi_{i})
\end{equation}

$\langle . \rangle$ denotes the temporal average.
$I_{c}(\phi)$ in Eq. \ref{general_form_input} therefore reads :

\begin{equation} \label{cosine_inhib_ring_input}
I_{c} (\phi) =  \sqrt{K} \left[I_{0c} + I_{\mu c} + I_{\mu c} \mu_{c}\cos 2\phi - \frac{J_{0}}{\pi} \int_{-\pi/2}^{\pi/2} \left[ 1 + 2p_{c} \cos 2 (\phi' - \phi) \right]\nu(\phi') \mathrm{d}\phi'\right]
\end{equation} 

In the large-K limit the equation above must be zero up to the $O(1/\sqrt{K})$ (balanced condition). Expressing the network firing rate based on its Fourier components, $\nu(\phi) = \nu_{0} + \sum_{i = 1}^{\infty} \nu_{i} \cos 2i\phi$, yields:

\begin{equation*}
\begin{split}
    I_{0c} + I_{\mu c} + I_{\mu c} \mu_{c}\cos 2\phi &= \frac{J_{0}}{\pi} \int_{-\pi/2}^{\pi/2} \left(1 + 2p_{c} \cos 2 (\phi' - \phi) \right)\left(\nu_{0} + \sum_{i = 1}^{\infty} \nu_{i} \cos 2i\phi'\right) \mathrm{d}\phi' \\
    &= \frac{J_{0}}{\pi} \left(
    \nu_{0}\pi + 2p_{c}\nu_{1}\int_{-\pi/2}^{\pi/2} \cos2(\phi' - \phi)\cos 2\phi'  \mathrm{d}\phi' \right
    )\\
    &= \frac{J_{0}}{\pi} \left(
    \nu_{0}\pi + 2p_{c}\nu_{1}\int_{-\pi/2}^{\pi/2} (\cos 2\phi' \cos 2\phi + \sin2\phi' \sin{2\phi'})\cos 2\phi'  \mathrm{d}\phi' \right) \\
    &= \frac{J_{0}}{\pi} \left(
    \nu_{0}\pi + 2p_{c}\nu_{1} \frac{\pi}{2} \cos{2\phi} \right) \\
\end{split}
\end{equation*} 

Therefore

\begin{equation}
\begin{split}
    I_{0c} + I_{\mu c} = J_{0} \nu_0 \\
    I_{\mu c}\mu_{c} = J_{0} p_{c}\nu_{1}
\end{split}
\end{equation} 

The average firing rate of the network reads:

\begin{equation}
\label{eq:cosine_response}
\boxed{
    \nu(\phi) = \frac{1}{J_{0}}(I_{0c} + I_{\mu c} (1 + \frac{\mu_{c}}{p_{c}}\cos2\phi)) + \sum_{i = 2}^{\infty} \nu_{i}\cos2i\phi}
\end{equation} 

From Eq.~\ref{cosine_inhib_ring_input}, $I_{c}(\phi)$ also has only two nonzero Fourier components:
\begin{equation} \label{averaged_input_two_comp}
I_{c}(\phi) = I_{0} + I_{1}\cos 2\phi.
\end{equation} 

\newpage
\begin{figure}[h!]
  \centering
  \includegraphics[width=\textwidth]{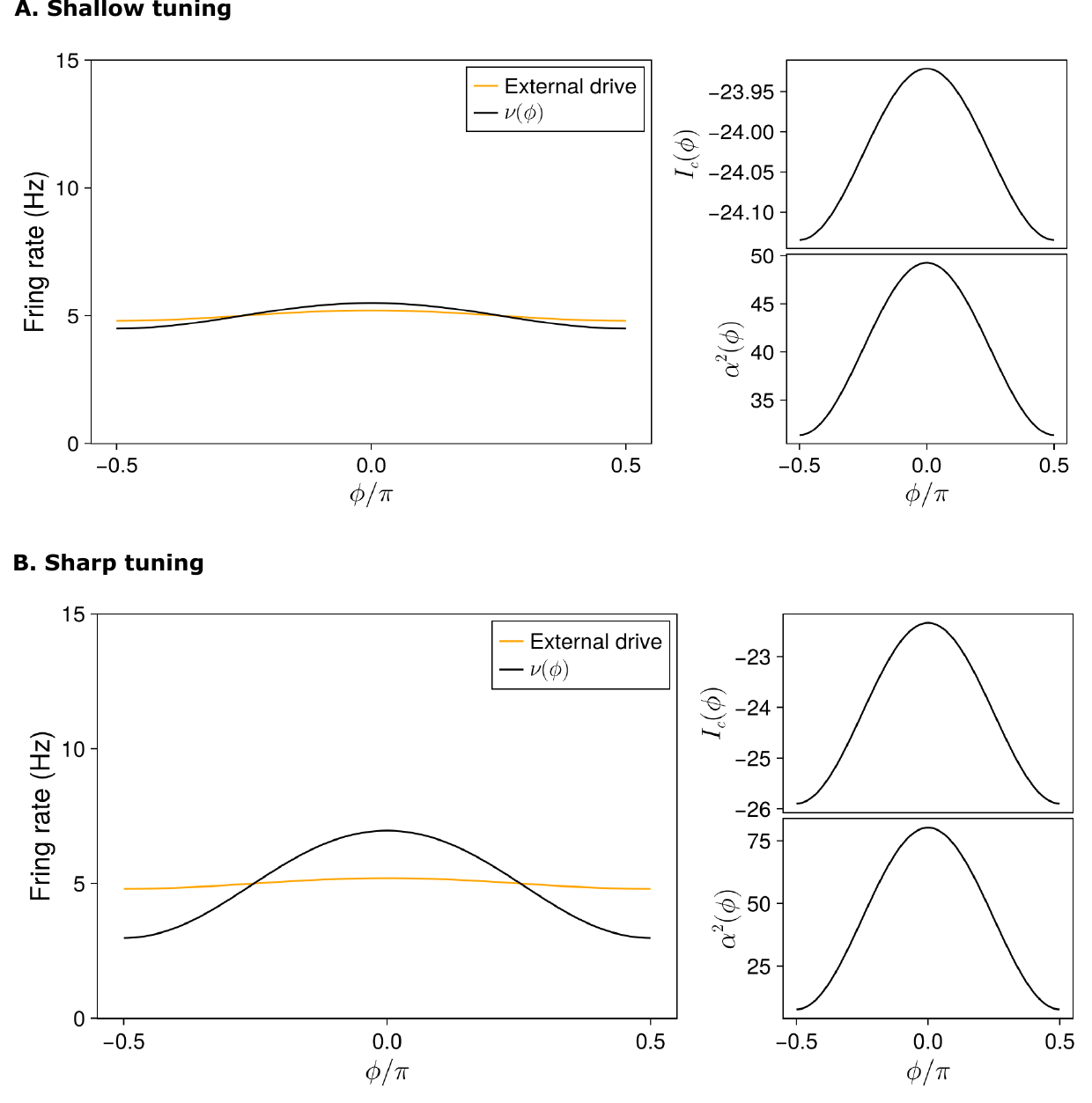} 
  \caption{\textbf{Shallow vs.\ sharp tuning in the cosine self-consistent network.} 
For both cases, the same external input is applied (orange), 
$ I_{\text{ext}}(\phi) = I_{0c} + I_{\mu c}\big(1 + \mu_c \cos(2\phi)\big) $, 
with $I_{0c}=1.0$, $I_{\mu c}=4.0$, $\mu_c=0.05$, $J_0=1.0$, $\tau_I=5\,\mathrm{ms}$, $\tau_M=10\,\mathrm{ms}$. 
The network output firing rate profile $\nu(\phi)$ (black) and the self-consistent recurrent mean input $I_c(\phi)$ 
and variance $\alpha^2(\phi)$ (right column) are shown. \textbf{(A)}~For narrower connection probability $p_c=0.4$ (more orientation-selective connectivity), the recurrent input weakly tunes the external drive, resulting in shallow tuning. 
\textbf{(B)}~For broader connection probability $p_c=0.1$, the untuned components of the stimulus are strongly suppressed, resulting in a sharp tuning profile.}
  \label{fig:cosine}
\end{figure}

Due to synaptic scaling, the temporal fluctuations become a smooth function of
location, so that $\sigma_{I,i}(\phi) \simeq \sigma_{I}(\phi)$. Considering Eq.~\ref{eq:exp_synapse_sigma}, \ref{eq:mean_varaince_input_rate_ring} we obtain
\begin{equation} \label{temporal_fluctuation}
\begin{split}
\sigma^{2}_{I}(\phi) &=  \frac{J_{0}^2}{2\tau_{I}K} 
\int_{-\pi/2}^{\pi/2} \omega(\phi - \phi') \nu(\phi') \,\mathrm{d}\phi' \\
&= \frac{J_{0}^2}{2\tau_{I}} \left(\nu_{0} + p_{c}\nu_{1}\cos 2\phi\right). \\
\end{split}
\end{equation} 

Since the balance equations determine $\nu_{0}$ and $\nu_{1}$, the temporal fluctuations of the input current depend only on the postsynaptic current kernel, not on the neuron model.
Using Gauss–Rice f–I curve Eq.~\ref{eq:fi_curve_gauss}, the temporal variance of the membrane potential is: 
\begin{equation}
\sigma_V^2 =\; \frac{J_0^2}{\tau_q} \left(\nu_{0} + p_{c}\nu_{1}\cos 2\phi\right)
\end{equation}

For the quenched fluctuations, using Eq.~\ref{eq:mean_varaince_input_rate_ring}, we have
\begin{equation} \label{quenched_fluctuation}
\begin{split}
\alpha^{2}(\phi) &= \left[ (I_{i}(\phi_{i}) - I_{c}(\phi))^2 \right] \\
&= \left[ (J\Omega_{i}(\phi_{i}) -  J\Omega(\phi))^2 \right] \\
&= J_{0}^{2} q(\phi) 
= J_{0}^{2}\int \omega(\phi' - \phi)\,\nu^2(\phi') \,\mathrm{d}\phi'.
\end{split}
\end{equation} 

Because $\omega(\Delta\phi)$ has only two nonzero Fourier components, $\alpha(\phi)$ can be written as
\begin{equation} \label{quenched_fluctuation_two_comp}
\alpha^{2}(\phi) = \alpha_{0}^{2} + \alpha_{1}^{2}\cos 2\phi
= J_{0}^{2}(q_{0} + q_{1}\cos 2\phi).
\end{equation} 

The following four self-consistent equations determine the four unknowns
$(I_{0}, I_{1}, \alpha_{0}, \alpha_{1})$, from which the mean population activity can be calculated:
\begin{equation}
\begin{split}
\nu_{0} &= \int_{-\pi/2}^{\pi/2} \nu(I_{i}) \,\rho(I_{i}) \,\mathrm{d}I_{i}\,\frac{\mathrm{d}\phi}{\pi}, \\
\nu_{1} &= \int_{-\pi/2}^{\pi/2} \nu(I_{i}) \,\rho(I_{i}) \,\mathrm{d}I_{i}\,\cos 2\phi\,\frac{\mathrm{d}\phi}{\pi}, \\
q_{0} &= \int_{-\pi/2}^{\pi/2} \nu^{2}(I_{i}) \,\rho(I_{i}) \,\mathrm{d}I_{i}\,\frac{\mathrm{d}\phi}{\pi}, \\
q_{1} &= \int_{-\pi/2}^{\pi/2} \nu^{2}(I_{i}) \,\rho(I_{i}) \,\mathrm{d}I_{i}\,\cos 2\phi\,\frac{\mathrm{d}\phi}{\pi}.
\end{split}
\end{equation}   
Based on Eq. \ref{general_form_input} the mean probability density of the input current reads:

\begin{equation}
\rho(I_{i}) = \frac{1}{\sqrt{2\pi}\alpha(\phi)} \exp \left(-\frac{\left(I_{i}(\phi_{i}) - I_{c}(\phi)\right)^2}{2\alpha^2(\phi)}\right) 
\end{equation}

$\nu(I_{i})$ is also determined using Gauss–Rice neuron \(f\)–\(I\) curve, considering also \ref{quenched_fluctuation_two_comp} and \ref{temporal_fluctuation}, after taking integral over $I_i$ we obtain:

\begin{equation}
\label{eq:self_consis_cosine}
\begin{split}
\nu_{0} &= \int_{-\pi/2}^{\pi/2} \frac{\nu_{m}J_{0}(\nu_{0}+ p_{c}\nu_{1}\cos2\phi)^{1/2}}{\tau_{q}^{1/2}\sqrt{\alpha^2_{0}+\alpha^2_{1}\cos{2\phi}+\frac{J_{0}^2}{\tau_{q}}(\nu_{0}+\nu_{1}p_{c}\cos{2\phi})}} \\
&\exp{\left(\frac{-(I_{0}+I_{1}\cos{2\phi}-\Psi_0)^2}{2(\alpha^2_{0}+\alpha^2_{1}\cos{2\phi}+\frac{J_{0}^2}{\tau_{q}}(\nu_{0}+\nu_{1}p_{c}\cos{2\phi}))}\right)}\frac{\mathrm{d}\phi}{\pi}\\
\nu_{1} &= \int_{-\pi/2}^{\pi/2} \frac{\nu_{m}J_{0}(\nu_{0}+ p_{c}\nu_{1}\cos2\phi)^{1/2}}{\tau_{q}^{1/2}\sqrt{\alpha^2_{0}+\alpha^2_{1}\cos{2\phi}+\frac{J_{0}^2}{\tau_{q}}(\nu_{0}+\nu_{1}p_{c}\cos{2\phi})}} \\
&\exp{\left(\frac{-(I_{0}+I_{1}\cos{2\phi}-\Psi_0)^2}{2(\alpha^2_{0}+\alpha^2_{1}\cos{2\phi}+\frac{J_{0}^2}{\tau_{q}}(\nu_{0}+\nu_{1}p_{c}\cos{2\phi}))}\right)}\frac{\cos{2\phi}\mathrm{d}\phi}{\pi}\\
\alpha^2_{0}/J^2_{0}  &= \int_{-\pi/2}^{\pi/2} \frac{\nu_{m}^2J_{0}(\nu_{0}+ p_{c}\nu_{1}\cos2\phi)^{1/2}}{\tau_{q}^{1/2}\sqrt{2(\alpha^2_{0}+\alpha^2_{1}\cos{2\phi})+\frac{J_{0}^2}{\tau_{q}}(\nu_{0}+\nu_{1}p_{c}\cos{2\phi})}} \\
&\exp{\left(\frac{-(I_{0}+I_{1}\cos{2\phi}-\Psi_0)^2}{2(\alpha^2_{0}+\alpha^2_{1}\cos{2\phi})+\frac{J_{0}^2}{\tau_{q}}(\nu_{0}+\nu_{1}p_{c}\cos{2\phi})}\right)}\frac{\mathrm{d}\phi}{\pi} \\
\alpha^2_{1}/J^2_{0} &= \int_{-\pi/2}^{\pi/2} \frac{\nu_{m}^2J_{0}(\nu_{0}+ p_{c}\nu_{1}\cos2\phi)^{1/2}}{\tau_{q}^{1/2}\sqrt{2(\alpha^2_{0}+\alpha^2_{1}\cos{2\phi})+\frac{J_{0}^2}{\tau_{q}}(\nu_{0}+\nu_{1}p_{c}\cos{2\phi})}} \\
&\exp{\left(\frac{-(I_{0}+I_{1}\cos{2\phi}-\Psi_0)^2}{2(\alpha^2_{0}+\alpha^2_{1}\cos{2\phi})+\frac{J_{0}^2}{\tau_{q}}(\nu_{0}+\nu_{1}p_{c}\cos{2\phi})}\right)}\frac{\cos{2\phi}\mathrm{d}\phi}{\pi}
\end{split}
\end{equation}

Numerically solving these self-consistency equations determines $I_0, I_1, \alpha_0, \alpha_1$. Having them $\nu(\phi)$ can be calculated using \ref{eq:self_mean_compact} (Fig. \ref{fig:cosine}): 

\begin{equation}
\boxed{
\begin{aligned}
\nu(\phi) &= \nu_m \frac{\sigma_V(\phi)}{\sqrt{\alpha^2(\phi)+\sigma_V(\phi)^2}}
\exp\!\left[-\frac{(I_c(\phi)-\Psi_0)^2}{2(\alpha^2(\phi)+\sigma_V(\phi)^2)}\right]\\
\alpha^{2}(\phi) &= \alpha_{0}^{2} + \alpha_{1}^{2}\cos 2\phi  \qquad I_c(\phi) = I_{0} + I_{1}\cos 2\phi \\
\sigma_V^2(\phi) & =\; \frac{J_0^2}{\tau_q} \left(\nu_{0} + p_{c}\nu_{1}\cos 2\phi\right) \\
\nu_0 &= \frac{I_{0c} + I_{\mu c}}{J_{0} }  \qquad \nu_{1} =  \frac{I_{\mu c}\mu_{c}}{J_{0} p_{c}} \\
\nu_m &= (2\pi\sqrt{\tau_I\tau_M})^{-1}
 \qquad  \tau_q = 2(\tau_I+\tau_M)
\end{aligned}}
\label{eq:self_mean_cosine}
\end{equation}

Using a second-order Taylor expansion, we analytically solve them for the weak modulation case i.e. $\alpha_{1}\ll \alpha_{0}$, $\nu_{1}\ll\nu_{0}$.

\begin{equation}
\begin{split}
\nu_{0} &= \frac{\nu_{m}J_{0}\nu_{0}^{1/2}}{\tau_{q}^{1/2}\sqrt{\alpha_{0}^2+\frac{J_{0}^2}{\tau_{q}}\nu_{0}}} \exp{\left(\frac{-(I_{0}-\Psi_0)^2}{2(\alpha_{0}^2+\frac{J_{0}^2}{\tau_{q}}\nu_{0})}\right)}\int_{-\pi/2}^{\pi/2} (1+a_{1}\cos{2\phi}+a_{0}\cos^2{2\phi})\frac{\mathrm{d}\phi}{\pi}\\
\nu_{1} &= \frac{\nu_{m}J_{0}\nu_{0}^{1/2}}{\tau_{q}^{1/2}\sqrt{\alpha_{0}^2+\frac{J_{0}^2}{\tau_{q}}\nu_{0}}} \exp{\left(\frac{-(I_{0}-\Psi_0)^2}{2(\alpha_{0}^2+\frac{J_{0}^2}{\tau_{q}}\nu_{0})}\right)}\int_{-\pi/2}^{\pi/2} (1+a_{1}\cos{2\phi}+a_{0}\cos^2{2\phi})\frac{\cos{2\phi}\mathrm{d}\phi}{\pi} \\
q_{0} &= \frac{\nu_{m}^2J_{0}\nu_{0}^{1/2}}{\tau_{q}^{1/2}\sqrt{2\alpha_{0}^2+\frac{J_{0}^2}{\tau_{q}}\nu_{0}}} \exp{\left(\frac{-(I_{0}-\Psi_0)^2}{2\alpha_{0}^2+\frac{J_{0}^2}{\tau_{q}}\nu_{0}}\right)}\int_{-\pi/2}^{\pi/2} (1+a'_{1}\cos{2\phi}+a'_{0}\cos^2{2\phi})\frac{\mathrm{d}\phi}{\pi} \\
q_{1}  &= \frac{\nu_{m}^2J_{0}\nu_{0}^{1/2}}{\tau_{q}^{1/2}\sqrt{2\alpha_{0}^2+\frac{J_{0}^2}{\tau_{q}}\nu_{0}}} \exp{\left(\frac{-(I_{0}-\Psi_0)^2}{2\alpha_{0}^2+\frac{J_{0}^2}{\tau_{q}}\nu_{0}}\right)}\int_{-\pi/2}^{\pi/2} (1+a'_{1}\cos{2\phi}+a'_{0}\cos^2{2\phi})\frac{\cos{2\phi}\mathrm{d}\phi}{\pi} 
\end{split}
\end{equation}   

Therefore we have:

\begin{equation}
\boxed{
\begin{split}
\nu_{0} &= \nu_{rnd}(1+\frac{a_{0}}{2}) \\
\nu_{1} &= \nu_{rnd}\frac{a_{1}}{2}\\
q_{0} &=  q_{rnd}(1+\frac{a'_{0}}{2})\\
q_{1}  &= q_{rnd}\frac{a'_{1}}{2}
\end{split}
}
\end{equation}

Here $\nu_{rnd}$ and $q_{rnd}$ correspond to the average network firing rate and quenched fluctuations, respectively, where the network connectivity is random i.e. previous section results (Eq. \ref{eq:self_mean_compact}, \ref{eq:self_second_compact}). $a_{0},a_{1},a'_{0},a'_{1}$ are coefficients ($\ll$1) that depend on parameters $\alpha_{0}, \alpha_{1}, I_{0}, I_{1}$.

\subsection{Von Mises modulated ring balanced networks}

We next consider a ring network with von--Mises modulated input and connectivity.  
The total synaptic input to neuron $i$ with preferred orientation $\phi_i$ reads  

\begin{equation} \label{total_input_von_mises}
  I_{i}(t,\phi_{i}) = \sqrt{K}\!\left(I_{0v} + I_{\mu v}\,\frac{e^{\kappa_{\mu}\cos 2(\phi_{0}-\phi_{i})}}{B_{0}(\kappa_{\mu})}\right) 
  + \sum_{j \in \mathrm{pre}(i)} J_{ij} f(t - t_{j}^{s}),
\end{equation}

where $\phi_{0}$ is the stimulus orientation, $\kappa_{\mu}$ sets the sharpness of external tuning, and $B_{0}$ is the modified Bessel function. The last term accounts for recurrent input.  

Connections are drawn with probability  

\begin{equation} \label{connection_p_von_mises}
  P_{ij} = \frac{K}{N}\,\frac{e^{\kappa_{p}\cos 2(\phi_{i}-\phi_{j})}}{B_{0}(\kappa_{p})},
\end{equation}

where $\kappa_{p}$ controls orientation specificity of recurrent coupling. 
Restricting to a single inhibitory population $(J_{ij}=-J_{0}/\sqrt{K})$, 
the analysis parallels the cosine case but with von--Mises kernels instead of cosines. 
The total input current again takes the form
\begin{equation*} \label{general_form_input_}
I_{i}(t, \phi_{i}) = I_{c}(\phi) + \alpha(\phi)x_{i} + \sigma_{I}(\phi)\,\eta(t),
\end{equation*}
but in the full--rank (von~Mises) setting, $I_{c}(\phi)$, $\alpha(\phi)$, and 
$\sigma_{I}(\phi)$ contain infinitely many Fourier components. Consequently, an 
infinite set of self--consistency equations would be required to determine them. 
This restricts the analysis to the balanced condition encoded in $I_{c}(\phi)$. 
A key advantage, however, is that from this condition the mean firing rate can 
be computed directly, since the von~Mises kernel allows the relevant integrals 
to be evaluated explicitly for all moments of the firing rate. Using Eq. \ref{average_recurernt_inhibitory}, $I_c$ reads:

\begin{equation} 
     I_c (\phi) =  \sqrt{K}\!\left(I_{0v} + I_{\mu v}  \frac{e ^ {\kappa_{\mu} \cos 2(\phi_0 - \phi)}}{B_{0}(\kappa_{\mu})}\right) 
     - \frac{J_{0}}{\sqrt{K}} \int_{-\pi/2}^{\pi/2} \! \omega(\phi' - \phi)\,\nu(\phi') \,\mathrm{d}\phi',
\end{equation} 
with kernel
\begin{equation}
\omega(\Delta\phi) = \frac{K}{\pi}\,\frac{e ^ {\kappa_{p} \cos (2 \Delta\phi)}}{B_{0}(\kappa_{p})}.
\end{equation}
Exploiting the symmetry of the network we set $\phi_{0} = 0$, which yields

\begin{equation}
\begin{split}
     I_c (\phi) =  &\sqrt{K} \left[I_{0v} + I_{\mu v} \frac{e ^ {\kappa_{\mu} \cos (2\phi)}}{ B_{0}(\kappa_{\mu})} - \frac{J_{0}}{\pi}\int_{-\pi/2}^{\pi/2} \frac{e ^ {\kappa_{p} \cos (2 (\phi - \phi'))}}{B_{0}(\kappa_{p})} \nu(\phi') \mathrm{d}\phi' \right].
\end{split}
\label{eq:von_input_averaged}
\end{equation} 

We use the expression of the von Mises cumulative distribution function as the series of modified Bessel function ratios \cite{mardia1975algorithm}:

\begin{equation} \label{com_von_dis}
\begin{split}
G(\phi) &= \int_{0}^{\phi} \frac{e^{\kappa \cos{z}}}{2\pi B_{0}(\kappa)} \mathrm{d}z \\
&= \frac{1}{2\pi B_{0}(\kappa)} \left(\phi B_{0}(\kappa) + 2\sum_{p=1}^{\infty}\frac{B_{p}(\kappa)\sin(p\phi)}{p}\right)
\end{split}
\end{equation} 

Taking the derivative of Eq. \ref{com_von_dis} yields:

\begin{equation}
\label{eq:von_expansion}
\frac{e ^ {\kappa \cos 2\phi}}{B_{0}(\kappa)} = 1 + 2 \sum_{i = 1}^{\infty} \frac{B_{i}(\kappa)}{B_{0}(\kappa)} \cos{2i\phi}
\end{equation} 

Using this expression the network averaged input Eq. \ref{eq:von_input_averaged} can be rewritten as:

\begin{equation} 
\label{eq:von_network_mean}
\begin{split}
     I_c (\phi) = &\sqrt{K} \left[I_{0v} + I_{\mu v} \left(1 + 2 \sum_{j = 1}^{\infty} \frac{B_{j}(\kappa_{\mu})}{B_{0}(\kappa_{\mu})} \cos{2j\phi} \right) \right. 
     \\ 
     & \left. \quad -\frac{J_{0}}{\pi}\int_{-\pi/2}^{\pi/2} \left(1 + 2 \sum_{k = 1}^{\infty} \frac{B_{k}(\kappa_{p})}{B_{0}(\kappa_{p})} \cos{2k(\phi-\phi')}\right) \nu(\phi') \mathrm{d}\phi'  \right].
\end{split}
\end{equation}

In the large-K limit the right hand side of Eq. \ref{eq:von_network_mean} must be zero up to the $O(1/\sqrt{K})$ (balanced condition). Considering this condition we obtain:

\begin{align*}
     & I_{0v} + I_{\mu v} \left[1 + 2 \sum_{j = 1}^{\infty} \frac{B_{j}(\kappa_{\mu})}{B_{0}(\kappa_{\mu})} \cos{2j\phi} \right]
      = J_{0} \left[\nu_{0} + \sum_{i = 1}^{\infty}  \nu_{i} \frac{B_{i}(\kappa_{p})}{B_{0}(\kappa_{p})}\cos{2i\phi}\right]
\end{align*}

Therefore:

\begin{equation}
\begin{split}
    \nu_{0} = & \frac{1}{J_{0}} (I_{0v} + I_{\mu v}) \\
    \nu_{i} = &\frac{2I_{\mu v}}{J_{0}} \frac{B_{i}(\kappa_{\mu})}{B_{i}(\kappa_{p})} \frac{B_{0}(\kappa_{p})}{B_{0}(\kappa_{\mu})}
\end{split}
\end{equation} 

The temporal average firing rate of the network in the general case of von Mises distribution is (Fig.~\ref{fig:vonmises_gaussian}):

\begin{equation}
\boxed{
    \nu(\phi) = \frac{1}{J_{0}}\left[I_{0} +  I_{\mu v}\left(1 + 2 \frac{B_{0}(\kappa_{p})}{B_{0}(\kappa_{\mu})} \sum_{i = 1}^{\infty}\frac{B_{i}(\kappa_{\mu})}{B_{i}(\kappa_{p})} \cos2i\phi\right)\right]
    }
\label{eq:von_complete_response}
\end{equation} 

\subsubsection{Gaussian approximation}
For the conditions where the modulation is strong, $\kappa \gg 1$, von Mises distribution can be approximated by a Gaussian distribution:

\begin{equation} \label{von_gaus_approximation}
     \frac{e ^ {\kappa \cos (2\phi)}}{ B_{0}(\kappa)} = \frac {\sqrt{2\pi}}{\sigma} e^{-
     \frac{(2\phi)^2}{2\sigma ^ 2} }
\end{equation} 

Where $\sigma ^ 2 = 1/\kappa$. Rewriting Eq. \ref{eq:von_input_averaged} and considering balanced condition we obtain:

\begin{equation} \label{inhibitory_von_balance_equation}
\begin{split}
     0 &=I_{0v} +I_{\mu v} \frac{e ^ {\kappa_{\mu} \cos (2\phi)}}{B_{0}(\kappa_{\mu})} - \frac{J_{0}}{\pi}\int_{-\pi/2}^{\pi/2} \frac{e ^ {\kappa_{p} \cos (2 (\phi - \phi'))}}{B_{0}(\kappa_{p})} \nu(\phi') \mathrm{d}\phi' \\
     &=I_{0v} + I_{\mu v} \frac {\sqrt{2\pi}}{\sigma_{\mu}} e^{-
     \frac{(2\phi)^2}{2\sigma_{\mu} ^ 2} } -  \frac{J_{0}}{\pi}\int_{-\pi/2}^{\pi/2} \frac {\sqrt{2\pi}}{\sigma_{p}} e^{-
     \frac{(2(\phi - \phi'))^2}{2\sigma_{p} ^ 2} } \nu(\phi') \mathrm{d}\phi'
\end{split}
\end{equation}

\newpage
\begin{figure}[h!]
  \centering
  \includegraphics[width=\textwidth]{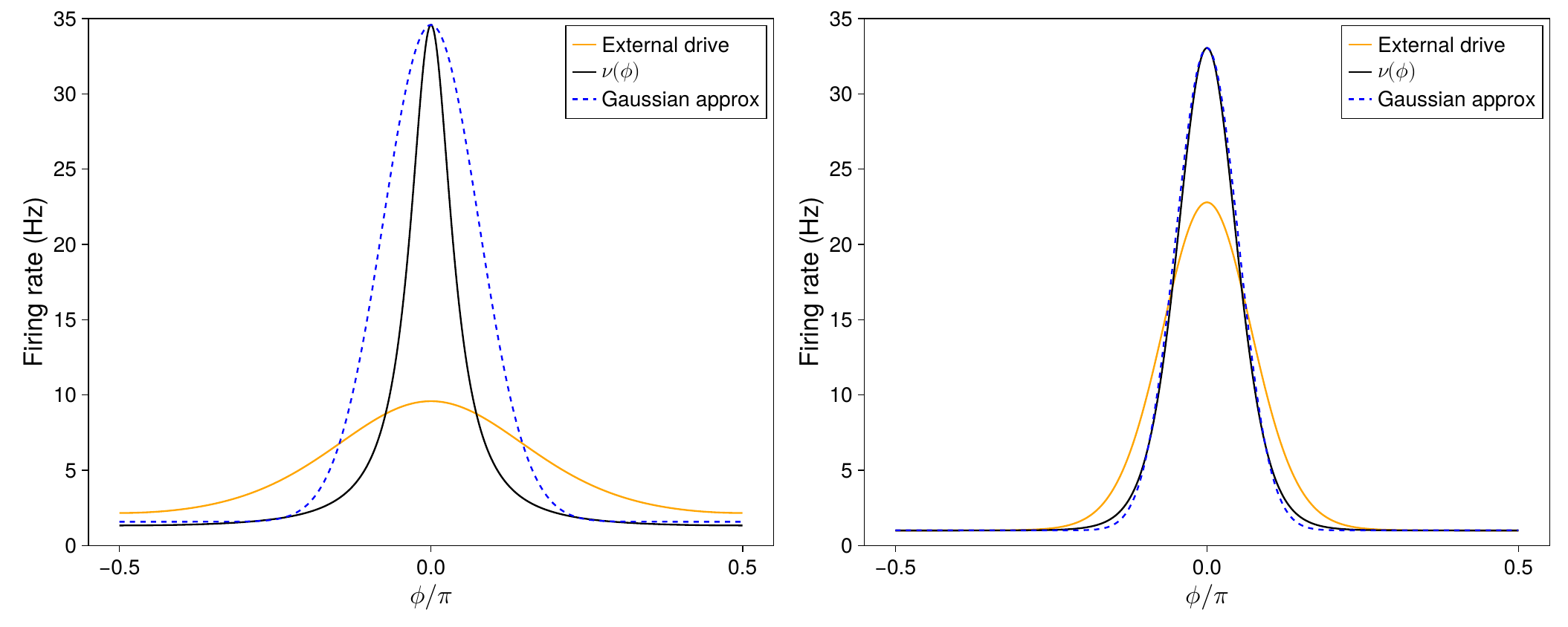} 
  \caption{\textbf{Firing rate profile of von--Mises modulated networks.} 
    Both panels show the network firing rate profile $\nu(\phi)$ (black) compared to the external drive (orange) and its Gaussian approximation (blue, dashed), for $I_{0v} = 1.0, \; I_{\mu v} = 4.0$. 
    The two cases correspond to different pairs of von--Mises concentration parameters: 
    \textbf{(Left)} $(\kappa_\mu, \kappa_p) = (1.0, 1.3)$ and 
    \textbf{(Right)} $(\kappa_\mu, \kappa_p) = (5.0, 10.0)$. 
    In the left panel, the Gaussian approximation fails to capture the profile accurately because $\kappa_\mu$ is small, and the von--Mises distribution is far from Gaussian. 
    Moreover, relative sharpening in the firing rate $\nu(\phi)$ is more pronounced when $\kappa_p$ is closer to $\kappa_\mu$, as seen in the left case.}
    \label{fig:vonmises_gaussian}
\end{figure}

Under Eq. \ref{von_gaus_approximation} approximation we also assume that the integral of the Gaussian over $-\pi,\pi$ is approximately equal to its integral over $-\infty,\infty$ limits. Based on the convolution theorem we can show that $\nu(\phi)$ has a Gaussian shape such that $\nu(\phi) = \nu_{0} + \nu_{1} \frac {\sqrt{2\pi}}{\sigma_{1}}  e^{-\frac{(2\phi)^2}{2\sigma_{1}^2}}$. We therefore can rewrite Eq. \ref{inhibitory_von_balance_equation}:

\begin{equation*}
\begin{split}
\label{inhibitory_von_balance_equation2}
     0 &=I_{0v} + I_{\mu v} \frac {\sqrt{2\pi}}{\sigma_{\mu}} e^{-
     \frac{(2\phi)^2}{2\sigma_{\mu} ^ 2} } - \frac{J_{0}}{\pi}
     \int_{-\pi/2}^{\pi/2} \frac {\sqrt{2\pi}}{\sigma_{p}} e^{-\frac{(2(\phi - \phi'))^2}{2\sigma_{p} ^ 2} } \left(\nu_{0} + \nu_{1}\frac {\sqrt{2\pi}}{\sigma_{1}} e^{-\frac{(2\phi')^2}{2\sigma_{1}^2}}\right) \mathrm{d}\phi'\\
     &= I_{0v} +  I_{\mu v} \frac {\sqrt{2\pi}}{\sigma_{\mu}} e^{-
     \frac{(2\phi)^2}{2\sigma_{\mu} ^ 2} } - \frac{J_{0}}{\pi}\left(\int_{-\pi/2}^{\pi/2} \nu_{0}\frac{\sqrt{2\pi}}{\sigma_{p}}e^{-\frac{(2(\phi - \phi'))^2}{2\sigma_{p} ^ 2}} \mathrm{d}\phi' + \int_{-\pi/2}^{\pi/2} \nu_{1}\frac {2\pi}{\sigma_{p}\sigma_{1}} e^{-\frac{(2(\phi - \phi'))^2}{2\sigma_{p} ^ 2} } e^{-\frac{(2\phi')^2}{2\sigma_{1}^2}}\mathrm{d}\phi'
     \right)\\
     &=I_{0v} + I_{\mu v} \frac {\sqrt{2\pi}}{\sigma_{\mu}} e^{-
     \frac{(2\phi)^2}{2\sigma_{\mu} ^ 2} } - \frac{J_{0}}{\pi} \left( \nu_{0}\pi + \frac{2\nu_{1}\pi^2 e^{-\frac{(2\phi)^2}{2(\sigma_{1}^{2}+ \sigma_{p}^{2})}}}{\sqrt{2\pi(\sigma_{1}^2 + \sigma_{p}^2)}} \right)
\end{split}
\end{equation*}

We obtain:
\begin{equation}
\begin{split}
    \nu_{0} &= \frac{I_{0 v}}{J_{0}} \\
    \nu_{1} &= \frac{I_{\mu v}}{J_{0}} \\
    \sigma_{1}^{2} &= \sigma_{\mu}^{2} - \sigma_{p}^{2}
\end{split}
\end{equation} 

Therefore the temporal average firing rate of the network reads (Fig.~\ref{fig:vonmises_gaussian}):

\begin{equation}
\label{eq:gaussian_limit_response}
\boxed{
\nu(\phi) = \frac{1}{J_{0}}\left[I_{0v} + I_{\mu v}\frac {\sqrt{2\pi}}{\sqrt{\sigma_{\mu}^{2} - \sigma_{p}^{2}}} e^{-\frac{(2\phi)^2}{2(\sigma_{\mu}^{2} - \sigma_{p}^{2})}}\right]
}
\end{equation} 

\subsubsection{Weak modulation}
In the case of small modulation $\kappa \ll 1$, a von Mises function can be well approximated by a Cosine function using it's Taylor expansion around $\kappa = 0$:

\begin{equation}
\label{eq:von_cosine_limit}
\frac{e ^ {\kappa \cos 2\phi}}{B_{0}(\kappa)} \approx \frac{ 1 + \kappa \cos{2\phi}}{B_{0}(\kappa)}
\end{equation} 

Using this approximation (and the balanced condition) Eq. \ref{eq:von_input_averaged} gives:

\begin{equation} \label{von_inhib_ring_input}
     I_{0v} + I_{\mu v} \frac{ 1 + \kappa_{\mu} \cos{2\phi}}{B_{0}(\kappa_{\mu})}  =   \frac{J_{0}}{\pi} \int_{-\pi/2}^{\pi/2} \frac{ 1 + \kappa_{p} \cos{2(\phi-\phi')}}{B_{0}(\kappa_{p})}\nu(\phi') \mathrm{d}\phi'.
\end{equation} 

In the large-K limit the equation above must be zero up to the $O(1/\sqrt{K})$ (balanced condition). Expressing the network firing rate based on its Fourier components, $\nu(\phi) = \nu_{0} + \sum_{i = 1}^{\infty} \nu_{i} \cos 2i\phi$, yields:

\begin{equation*}
\begin{split}
    I_{0v} + I_{\mu v} \frac{ 1 + \kappa_{\mu} \cos{2\phi}}{B_{0}(\kappa_{\mu})} &= \frac{J_{0}}{\pi} \int_{-\pi/2}^{\pi/2} \left( \frac{ 1 + \kappa_{p} \cos{2(\phi-\phi')}}{B_{0}(\kappa_{p})}\right)\left(\nu_{0} + \sum_{i = 1}^{\infty} \nu_{i} \cos 2i\phi'\right) \mathrm{d}\phi' \\
    &= \frac{J_{0}}{B_{0}(\kappa_{p})\pi} \left(
    \nu_{0}\pi + \kappa_{p}\nu_{1}\int_{-\pi/2}^{\pi/2} \cos2(\phi' - \phi)\cos 2\phi'  \mathrm{d}\phi' \right
    )\\
    &= \frac{J_{0}}{B_{0}(\kappa_{p})\pi} \left(
    \nu_{0}\pi + \kappa_{p}\nu_{1}\int_{-\pi/2}^{\pi/2} (\cos 2\phi' \cos 2\phi + \sin2\phi' \sin{2\phi'})\cos 2\phi'  \mathrm{d}\phi' \right) \\
    &= \frac{J_{0}}{B_{0}(\kappa_{p})\pi} \left(
    \nu_{0}\pi + \kappa_{p}\nu_{1} \frac{\pi}{2} \cos{2\phi} \right) \\
\end{split}
\end{equation*} 

Therefore

\begin{equation}
\begin{split}
   I_{0v} + \frac{I_{\mu v}}{B_{0}(\kappa_{\mu})} = \frac{J_{0} \nu_0}{B_{0}(\kappa_{p})} \\
     \frac{I_{\mu v}\kappa_{\mu}}{B_{0}(\kappa_{\mu})} =  \frac{J_{0}\kappa_{p}\nu_1}{2B_{0}(\kappa_{p})} 
\end{split}
\end{equation} 

The average firing rate of the network reads:

\begin{equation} \label{weak_modulation_von_mises}
\boxed{
    \nu(\phi) = \frac{B_{0}(\kappa_{p})}{J_{0}}\left[ I_{0v} +  \frac{I_{\mu v}}{B_{0}(\kappa_{\mu})} \left(1 + \frac{2\kappa_{\mu}}{\kappa_{p}}cos2\phi\right)\right] + \sum_{i = 2}^{\infty} \nu_{i}cos2i\phi
    }
\end{equation} 

This solution is similar to the cosine modulated scenario. One can see the networks are approximately the same under the conditions:

\begin{equation}
\label{eq:parameter_transition}
\begin{split}
    I_{0c} = I_{0v} \\
    I_{\mu c} = I_{\mu v} \\
    2p_{c} = \kappa_{p} \\
    \mu_{c} = \kappa_{\mu}
\end{split}
\end{equation}

\subsection{Universality of the population profile}

We addressed the universality of the mean population profile in balanced structured networks through a comparative approach. We analyzed high-rank (von~Mises) structured networks. Using the input average (Eq.~\ref{eq:von_input_averaged}), the balance condition (requiring inputs to remain $\mathcal{O}(1)$ in the large-$K$ limit), and the von~Mises expansion (Eq.~\ref{eq:von_expansion}), we derived a closed-form expression for the population firing-rate profile as an infinite Bessel series (Eq.~\ref{eq:von_complete_response}). This expression remains valid across tuning regimes, from high modulation (Gaussian limit; Eqs.~\ref{von_gaus_approximation}, \ref{eq:gaussian_limit_response}) to low modulation (cosine limit; Eq.~\ref{eq:von_cosine_limit}), and is independent of both the neuronal transfer function and the synaptic kernel. Thus, in the von~Mises case the balance condition alone uniquely determines the mean firing-rate profile.

By contrast, in low-rank (cosine) structured networks the balance condition (Eq.~\ref{cosine_inhib_ring_input}) constrains only the first two Fourier components of the firing rate, $\nu_{0}$ and $\nu_{1}$ (Eq.~\ref{eq:cosine_response}), while higher-order components remain undetermined. To resolve them, one must specify a concrete neuronal transfer function, such as the Gauss–Rice model (Eq.~\ref{eq:fi_curve_gauss}), together with the synaptic kernel, and solve the resulting self-consistency equations (Eq.~\ref{eq:self_consis_cosine}). In contrast to the von~Mises case, the cosine network therefore requires explicit neuronal dynamics to reconstruct the full population profile.

\begin{figure}[h!]
    \centering
    \includegraphics[width=\textwidth]{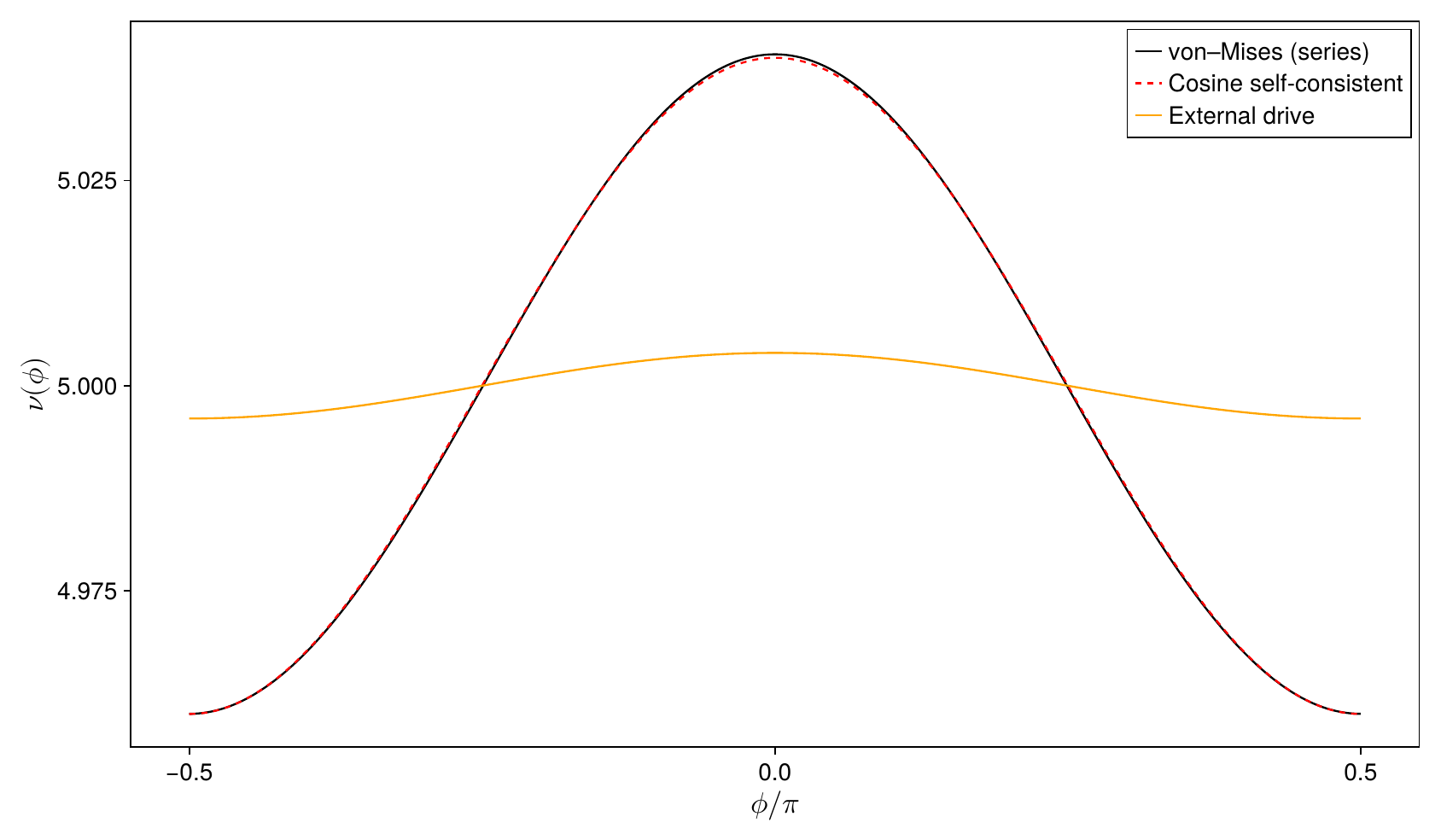}
    \\
    \includegraphics[width=\textwidth]{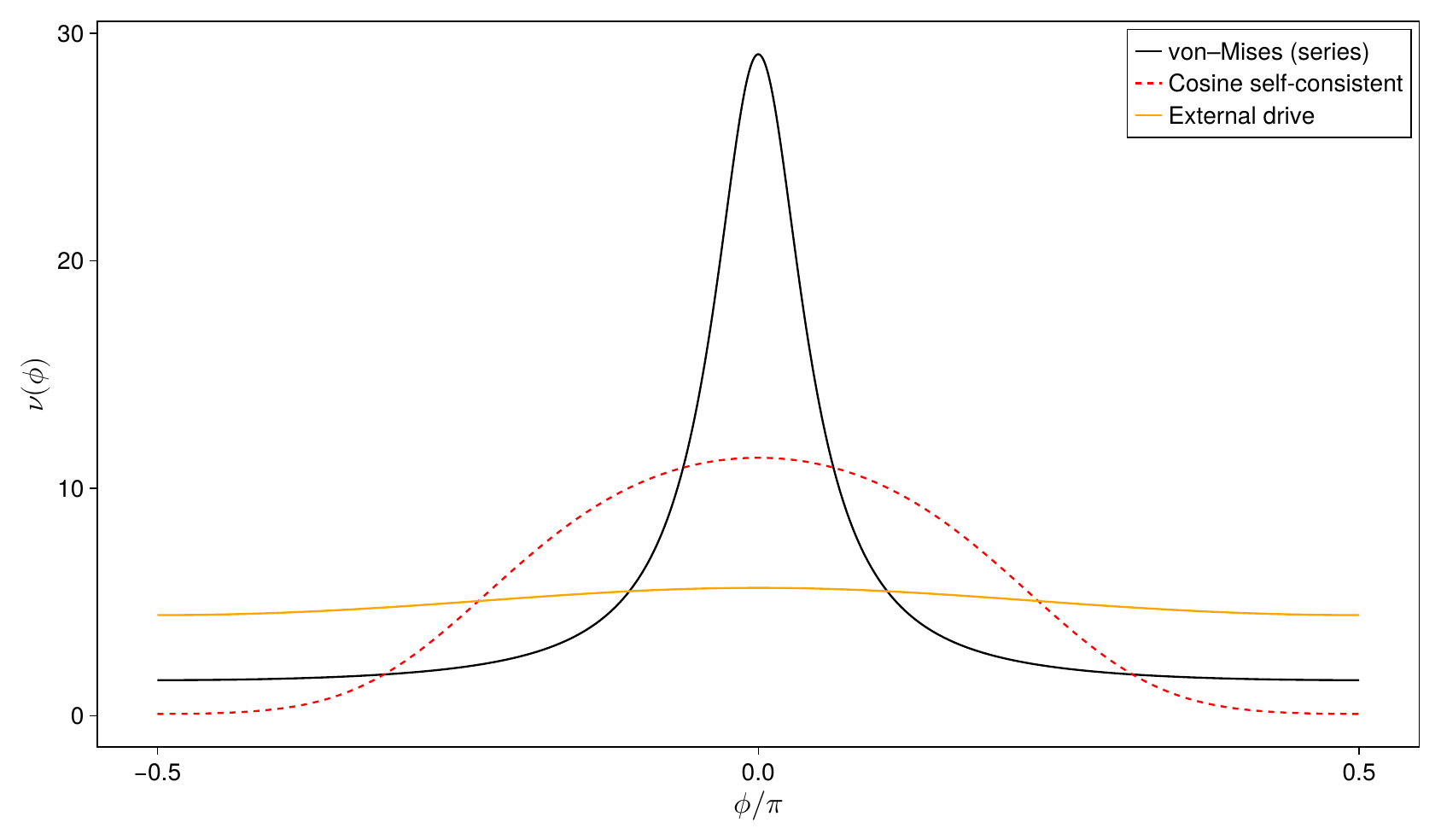}
    \caption{\textbf{Comparison of network tunings.} 
    (Top) Within the universality regime, where the von--Mises kernel is well captured by its first Fourier (cosine) component and the resulting solution coincides with the self-consistent prediction. 
    In this case, both input and connectivity are only weakly modulated, and sharpening is shallow, i.e. the orientation selectivity of the input is much broader than that of the connectivity. 
    (Bottom) Outside the universality regime, where the orientation selectivity of the input approaches that of the connectivity. 
    Here, higher Fourier components of the von--Mises kernel contribute significantly, and the cosine approximation no longer reproduces the self-consistent solution. 
    Parameters: $I_{0v} = 1.0$, $I_{\mu v} = 4.0$, $\kappa_p = 0.2$, with $\kappa_\mu = 0.001$ (top) and $\kappa_\mu = 0.15$ (bottom).}
    \label{fig:universality}
\end{figure}

Finally, we showed that in the weak-modulation limit the von~Mises kernel reduces to a cosine (Eq.~\ref{eq:parameter_transition}, Fig.~\ref{fig:universality}). In this regime, the von~Mises analysis based solely on the balance condition reproduces the cosine result, which otherwise demands solving the self-consistency equations. It should be noted that when input selectivity approaches the connection selectivity ($\kappa_{\mu} \to \kappa_{p}$), the cosine approximation is no longer valid because higher Fourier components of the von~Mises kernel contribute significantly. In this case, the shape of the population profile under von~Mises connectivity differs from that of the cosine approximation (Fig.~\ref{fig:universality}).

\newpage

\subsection{Heterogeneity in balanced ring networks}

We examined how heterogeneity is captured in different network architectures. For
high rank connectivity, the balance condition determines only the mean firing-rate profile
(Eq.~\ref{eq:von_complete_response}) but does not constrain variability across neurons.
Accounting for heterogeneity in this case would require solving an infinite set of
self-consistency equations, one pair for each Fourier mode of the firing-rate distribution, which is
analytically intractable.

In contrast, for low rank connectivity the balance condition alone is insufficient to determine the
full profile, but the corresponding self-consistency analysis reduces to four variables,
$(I_{0}, I_{1}, \alpha_{0}, \alpha_{1})$ (Eq.~\ref{eq:self_consis_cosine}). These jointly determine the
mean response $\nu(\phi)$ (Eq.~\ref{eq:self_mean_cosine}) and the quenched heterogeneity,
with $\alpha_{0}^{2}=J_{0}^{2}q_{0}$ and $\alpha_{1}^{2}=J_{0}^{2}q_{1}$ (Eq.~\ref{quenched_fluctuation_two_comp}, Fig.~\ref{fig:cosine}). Thus, the cosine case provides a tractable
framework that captures both the average population activity and its across-neuron variability.

\subsection{Response sharpening through inhibition}

Our framework also explains how inhibitory feedback shapes input selectivity. Within the von Mises framework, the balance condition yields a closed-form firing-rate profile (Eq.~\ref{eq:von_complete_response}, Fig. \ref{fig:vonmises_gaussian}). From this expression we find that the ratio of output to input modulation depends only on the tuning parameters $\kappa_{\mu}$ and $\kappa_{p}$, implying contrast invariance: inputs of different amplitudes but identical tuning are sharpened by the same factor.

The Bessel coefficients decrease with harmonic order $i$, and their decay rate is controlled by the relative tuning widths. Sharpening increases as the input and connectivity tunings become matched, i.e.\ as $\kappa_{\mu}\!\to\!\kappa_{p}$ from below, because the harmonic decay becomes slower and higher modes contribute more strongly (Eq.~\ref{eq:von_complete_response}, Fig. \ref{fig:vonmises_gaussian}). When the input is much broader than the connectivity ($\kappa_{\mu}\ll\kappa_{p}$), sharpening is weak (Eq.~\ref{eq:von_complete_response}, Fig. \ref{fig:vonmises_gaussian}). If the input is narrower than the connectivity ($\kappa_{\mu} > \kappa_{p}$), the balanced solution is not well defined, and no consistent profile is obtained.

\section{Discussion}
In this study we presented, for the first time, a mean-field approach to spiking balanced ring networks that is analytically tractable. Using a low-rank structured network, we showed how biophysical details of the neuron model together with the synaptic kernel shape heterogeneity across neurons, providing a tractable framework to capture variability that is inaccessible in the high-rank setting. For the high-rank case with von~Mises connectivity, we derived a closed-form solution for the mean population profile. This analysis revealed a universal structure: the population response is uniquely fixed by the balance condition, independent of neuronal transfer functions or synaptic kernels. Finally, we demonstrated how, by varying the network modulation, the von~Mises and cosine descriptions align in the weak-modulation limit.

Our closed-form solution further demonstrated how inhibition gives rise to response sharpening and contrast invariance, both of which are characteristic properties of cortical dynamics. In this framework, inhibitory feedback narrows tuning curves in a way that depends only on the relative modulation of external input and recurrent connectivity, thereby ensuring contrast-invariant selectivity. This sharpening mechanism is distinct from excitation-driven sharpening, as it relies on inhibitory balance rather than enhanced excitatory drive, highlighting a complementary pathway by which cortical networks can generate selective responses.

Our framework is readily generalizable. By allowing for a broad class of synaptic kernels, the same mean-field analysis can capture diverse forms of structured connectivity. Likewise, retaining additional Fourier components in the self-consistency equations enables the description of more complex phenomena, such as direction selectivity, which requires keeping the first three components of the mean field. In this way, the approach provides a flexible foundation for extending balanced ring models to a wide range of computational functions observed in cortical circuits.

An interesting direction for future work is to incorporate short-term synaptic plasticity into the present framework, in particular to account for recurrent excitation. Such extensions would allow us to study how transient synaptic dynamics interact with inhibitory balance in shaping tuning. Indeed, Mongillo and colleagues have already developed a mean-field description of short-term plasticity in random networks and showed that it can generate bistable solutions \cite{mongillo2012bistability}. Embedding similar mechanisms into balanced ring models may reveal how short-term dynamics interact with structured inhibition to modulate selectivity. A second promising avenue is to extend the treatment of heterogeneity. While here we focused on mean responses and their variance, the mathematical framework developed by van~Vreeswijk and Sompolinsky \cite{van2005irregular} shows how correlations of quenched disorder across orientations can be systematically analyzed. Incorporating such methods would make it possible to capture tuning-curve heterogeneity at the level of individual neurons, thereby bringing theory into even closer contact with experimental variability in cortical circuits.

\section{Acknowledgements}

This work was funded by the Deutsche Forschungsgemeinschaft (DFG, German Research Foundation) - Project-ID 454648639 - SFB 1528 – A01 $\&$ C01; Project-ID 436260547 in relation to NeuroNex (National Science Foundation  2015276); Project-ID 430156276 - as part of the SPP 2205; and by the Niedersächsisches Vorab of the VolkswagenStiftung through the Göttingen Campus Institute for Dynamics of Biological Networks (ZN3326 $\&$ ZN3371).
\clearpage

\bibliographystyle{unsrt}
\bibliography{references}
\end{document}